\documentstyle[preprint,aps,floats]{revtex}
\input epsf.tex
\def\DESepsf(#1 width #2){\epsfxsize=#2 \epsfbox{#1}}
\begin{document}
\preprint{\vbox{\hbox{}}} \draft

\title{
Constraints on Supersymmetric Flavor Changing Parameters\\ Using $B\to PP$ Decays}%
\vfill
\author{Dilip Kumar Ghosh, Xiao-Gang He, Yu-Kuo Hsiao and Jian-Qing Shi}
\address{
\rm Department of Physics, National Taiwan University, Taipei,
Taiwan 10764, R.O.C.}

%\date{\today}
%
%\vskip -1cm
%
\vfill
\maketitle
\begin{abstract}
We study contributions of quark-squark-gluino interactions in
Minimal Supersymmetric Standard Model (MSSM) to $B\to PP$ ($PP =K
\pi, \pi\pi, KK$) decays using QCD improved factorization method
for the evaluation of the hadronic matrix elements and taking into
account renormalization group running of the Wilson coefficients
from SUSY scale ($\approx m_{\tilde q} $) to the low energy scale
($= m_b$) applicable to $B$ decays. Using the most recent
experimental data we obtain constraints on flavor changing
Supersymmetric (SUSY) parameters $(\delta_{ij})_{LL, RR}$. For
$\Delta S = -1$ processes, $b\to s \gamma$ is usually considered
to give the strongest limits. We, however, find that in some part
of the parameter space $B\to K \pi$ processes give stronger
bounds. Implications for $B^0_s -\bar B^0_s$ mixing is discussed.
We also study constraints obtained from $\Delta S = 0$ processes
$B\to \pi\pi, KK, \rho \gamma$ and $B_d -\bar B_d$ mixing. In this
case, in a large part of the parameter space $B_d - \bar B_d$
provides the best bound, but $B\to K^- K^0, \rho \gamma$ can still
give interesting constraints in a complementary region of the
parameter space.

\end{abstract}

\newpage

\section{Introduction}

Recent results from  $B$-factories at CLEO, BaBar,
Belle have attracted a lot of attentions. Data from these experiments
provide new opportunities to further study the Standard Model (SM) and
also to study physics beyond the SM.
Rare charmless hadronic $B$ decays, such as
$B\to PP (PP =K \pi, \pi\pi, KK)$, are particularly interesting.
These processes being rare are sensitive to new physics beyond
the SM. Minimal Supersymmetric Standard Model (MSSM),
which is an extension of the SM emerges as one of the most promising
candidates for new physics beyond the SM. MSSM resolve many of the potential
problem of the SM, for
example the hierarchy problem, unification of
$SU(3)\times SU(2)\times U(1)$ gauge couplings, and so on \cite{Haber}.

$B\to PP$ decays have been studied extensively within the framework
of naive factorization scheme. In the last few years, QCD improved factorization
method have been developed.
This method incorporates elements of the naive
factorization approach (as its leading term) and perturbative QCD
corrections (as subleading contributions) allowing one to compute
systematic radiative corrections to the naive factorization for the
hadronic $B$ decays~\cite{BBNS,hn}.
This method provides a better understanding of
the strong interaction dynamics for $B$
decay into two light mesons. Therefore results obtained using QCD
improved method are expected to be more reliable compared with those
obtained by using naive factorization. Lot of studies have been done
in this direction so far \cite{DDG,he1}.

In this paper we study implications of the recently measured
$B\to PP$ branching ratios on SUSY flavor changing parameters,
specially arising from the
quark-squark-gluino
($q-\tilde q - \tilde g$) interactions
in the model\cite{FCNC}. Effects of SUSY interactions
on various processes involving $B$ mesons have been
studied previously\cite{FCNC,btopp,bdmix,bsg,BBM,Branco}. Here we
will concentrate on the $q-\tilde q - \tilde g$
interactions and re-analyze $B\to PP$
decays with several improvements. Firstly we use the
QCD improved factorization method to compute the two body hadronic $B$ decays.
Secondly, we improve the quark level effective
Hamiltonian by including QCD running of different Wilson coefficients (WC)
from the SUSY scale ($=m_{\tilde q} $) to the low energy scale ($=m_b $)
which is applicable to $B$ decays. Previous authors while doing similar
analysis have neglected this RG evolution of WC's.
We find that this RG evolution of WC's have
non-negligible effects on the branching ratio calculations.

Using the most recent experimental data from different $B$-factories on
the $B \to PP $ decays we constrain the flavor changing
SUSY parameters arising from the $\tilde q_{iLR}-\tilde q_{jLR}$ mixing
due to the presence of $q-\tilde q -\tilde g $
interaction vertices in different one loop contributions to $B \to PP$ decays.
Similar kind of squark mixing can also occur for other
flavor changing $B$ decays. For $\Delta S = -1$ processes,
$b\to s \gamma$
is usually considered to give the strongest limit on
such mixing. We, however, find that
in some part of the parameter space $B\to K \pi$ processes
provide better bound.  We also discuss the implications of $B^0_s -\bar B^0_s$
mixing in this context. Furthermore, we also study constraints on FCNC
parameters obtained from $\Delta S = 0$ processes
$B\to \pi\pi, KK, \rho \gamma$,
and $B_b - \bar B_d$ mixing. We find that $B^0_d -\bar B^0_d$ mixing provides
the best limit in a large part of the parameter space, but in certain region of
the parameter space $B\to \pi\pi, KK, \rho\gamma$ can still give better
constraints.

\section{Squark-gluino contributions to $B\to PP$ Decays}

In the SUSY, squark and gluino induce large one loop contribution to
$B \to PP$ decay from penguin and box diagrams
because they are related to strong couplings.
We compute the effect of flavor changing contributions to $B \to PP $
arising from $q-\tilde q -\tilde g$ interaction vertices using
the mass insertion approximation method \cite{FCNC,Hall}.
In this method, the basis of the quark and squark states are chosen in such
a way that their couplings with gluinos are flavor
diagonal. The flavor changing current arises from the non-diagonality of the
squark propagators. We denote the off-diagonal terms in the squark mass
matrices (i.e. the mass terms relating squark of the same electric charge,
but different flavor), by $\Delta $, then we expand the squark propagators
as series in terms of $(\delta_{ij})_{AB}=(\Delta_{ij})_{AB}/{\tilde m}^2$
where ${\tilde m}$ is an average squark mass, $A, B = L, R $ and $i,j$ are the
generation indices. The notation $(\delta_{13})_{AB}$ and $(\delta_{23})_{AB}$
denote ${\tilde d}-{\tilde b}$ and ${\tilde s}-{\tilde b}$ mixing respectively.
These two parameters signify the size of flavor changing interactions
in SUSY through the $q-\tilde q-\tilde g$ interactions.

At one loop level, for $\tilde q_L -\tilde q_L$ mixing ($LL$ type ), the
new effective Hamiltonian relevant to $B\to K \pi, \pi \pi$
decays arise from SUSY penguin and box diagrams with gluino-squark in the
loops is given by
\begin{eqnarray}
H_{eff}^{susy} =-{G_F\over \sqrt{2}} V_{tb} V_{tq}^*
\sum_{i=3}^{6}C_i^{LL} O_i.
\end{eqnarray}
Here we have normalized the WC's
according to the SM ones.
The operator $O_i$ is the same as SM definition with
\begin{eqnarray}
&&O_{3} = \bar q_i\gamma^\mu L b_i\sum_{q'}
\bar q^\prime_j \gamma_\mu L q^\prime_j,\;\;
O_{4} = \bar q_i\gamma^\mu L b_j\sum_{q'}
\bar q^\prime_j\gamma_\mu L q^\prime_i,\nonumber\\
&&O_{5} = \bar q_i\gamma^\mu L b_i\sum_{q'}
\bar q^\prime_j \gamma_\mu R q^\prime_j,\;\;
O_{6} = \bar q_i\gamma^\mu L b_j\sum_{q'}
\bar q^\prime_j\gamma_\mu R q^\prime_i,
\label{LL}
\end{eqnarray}
where $L(R)=(1\pm\gamma_5)$, $q=d,s$, $q^\prime = u,d,s$.
The WC's are given by \cite{FCNC}:
\begin{eqnarray}
C_3^{LL}(M_{susy}) &=& \frac{-\alpha_s^2}{2 \sqrt{2} G_F  V_{tb}
V_{tq}^* m_{\tilde{q}}^2}\left(\delta_{j3}\right)_{LL} \left(  -
\frac{1}{9} {\mbox B}_1(x) - \frac{5}{9}{\mbox B}_2(x) -
\frac{1}{18}{\mbox P}_1(x) - \frac{1}{2}{\mbox P}_2(x)\right)\;
\nonumber  \\ C_4^{LL}(M_{susy}) &=& \frac{-\alpha_s^2}{2 \sqrt{2}
G_F V_{tb} V_{tq}^* m_{\tilde{q}}^2}\left(\delta_{j3}\right)_{LL}
\left(  - \frac{7}{3} {\mbox B}_1(x) + \frac{1}{3}{\mbox B}_2(x) +
\frac{1}{6}{\mbox P}_1(x) + \frac{3}{2}{\mbox P}_2(x)\right)\;
\nonumber  \\ C_5^{LL}(M_{susy}) &=& \frac{-\alpha_s^2}{2 \sqrt{2}
G_F V_{tb} V_{tq}^* m_{\tilde{q}}^2}\left(\delta_{j3}\right)_{LL}
\left( \frac{10}{9} {\mbox B}_1(x) + \frac{1}{18}{\mbox B}_2(x) -
\frac{1}{18}{\mbox P}_1(x) - \frac{1}{2}{\mbox P}_2(x)\right)\;
\nonumber  \\ C_6^{LL}(M_{susy}) &=& \frac{-\alpha_s^2}{2 \sqrt{2}
G_F V_{tb} V_{tq}^* m_{\tilde{q}}^2}\left(\delta_{j3}\right)_{LL}
\left(  - \frac{2}{3} {\mbox B}_1(x) + \frac{7}{6}{\mbox B}_2(x) +
\frac{1}{6}{\mbox P}_1(x) + \frac{3}{2}{\mbox P}_2(x)\right)\;
\label{CLL}
\end{eqnarray}
where $j=1,2$ when $q=d,s$, respectively and
$x = m^2_{\tilde g}/m^2_{\tilde q}$ . The definition of the
functions $B_i(x), P_i(x)$ can be found in Ref\cite{FCNC}.

The above WC's are obtained at the SUSY scale $M_{susy}$.
We take this scale as $m_{\tilde q}$. We then evolve these WC's from this
high scale down to the scale $m_b$ using the RG equation.
At the lower scale, after the RG evolution, the WC's take following form:
\begin{eqnarray}
C_3^{LL}(m_b)&=&b_3^3 C_3^{LL}(M_{susy})+b_3^4 C_4^{LL}(M_{susy})
+b_3^5 C_5^{LL}(M_{susy})+b_3^6 C_6^{LL}(M_{susy})\nonumber \\
C_4^{LL}(m_b)&=&b_4^3 C_3^{LL}(M_{susy})+b_4^4 C_4^{LL}(M_{susy})
+b_4^5 C_5^{LL}(M_{susy})+b_4^6 C_6^{LL}(M_{susy})\nonumber \\
C_5^{LL}(m_b)&=&b_5^3 C_3^{LL}(M_{susy})+b_5^4 C_4^{LL}(M_{susy})
+b_5^5 C_5^{LL}(M_{susy})+b_5^6 C_6^{LL}(M_{susy})\nonumber \\
C_6^{LL}(m_b)&=&b_6^3 C_3^{LL}(M_{susy})+b_6^4 C_4^{LL}(M_{susy})
+b_6^5 C_5^{LL}(M_{susy})+b_6^6 C_6^{LL}(M_{susy})
\end{eqnarray}
$b^j_i$ are the constant after running, depend on
the choice of the initial scale, as shown in Table. \ref{susywilson}.
\begin{table}[htb]
\begin{tabular}{|c|c|c|c|c|c|c|c|c|}
\hline $M_{susy}~(\rm GeV)$ & $b^3_3$ & $b^4_3$ & $b^5_3$ &
$b^6_3$ & $b^3_4$ & $b^4_4$ & $b^5_4$ &  $b^6_4$ \\
\hline 100& 1.144& -0.222& 0.009& 0.086&-0.322&0.989&-0.022&-0.197\\
\hline 500& 1.208& -0.299& 0.018& 0.139&-0.432&1.002&-0.042&-0.298\\
\hline 1000& 1.235& -0.328& 0.022& 0.164&-0.474&1.009&-0.051&-0.342\\
\hline
\hline $M_{susy}~(\rm GeV)$ & $b^3_5$ & $b^4_5$ & $b^5_5$ &
$b^6_5$ & $b^3_6$ & $b^4_6$ & $b^5_6$ &  $b^6_6$ \\
\hline 100& 0.011& 0.039& 0.925& 0.058& -0.051&-0.173&0.330&1.742\\
\hline 500& 0.011& 0.050& 0.905& 0.083&-0.064&-0.258&0.478&2.084\\
\hline 1000& 0.011& 0.053& 0.898& 0.094&-0.068&-0.294&0.540&2.227\\
\hline
\end{tabular}
\smallskip

\caption{Values of different WC's after RG evolution for three
 values of $M_{susy}=100, 500$ and $1000$~GeV.} \label{susywilson}
 \end{table}

We find that the WC's with and without the RG evolution
can be different by a factor of two in some cases. This implies
that to have a reliable result and consistent computation one has to take into
account the RG evolution of WC's.

Apart from $LL$ type of operators, one can also have operators with
different chiral structures, such as $RR, LR $ and $RL$ type.
The operators and WC's for $RR $ mixing
can be obtained by the exchange $L \leftrightarrow R$ in Eqs.
\ref{LL} and \ref{CLL}. $LR$ and $RL$ mixing cases can also be easily studied.
However, in these two cases there are more stringent constraints from
gluonic dipole moment operators for $B\to PP$ decays. The details
have been studied in \cite{He_self}. We will not discuss these two cases
further.

\section{Decay amplitudes from QCD Improved Factorization}

To obtain the branching ratios of $B\to PP$ decays,
one needs to study the matrix elements,
$<PP|H_{eff}|B>$. The usual approach is the naive factorization
method, where the hadronic matrix element can be approximated as a product
of two single current matrix elements, then it is parametrized into
meson decay constant and meson-meson transition form factor.

Previous analysis of SUSY contribution to $B \to PP$ decays
have been performed using the above mentioned naive factorization method.
However, in the present analysis we will use the QCD improved
factorization approach\cite{BBNS} to
compute the SUSY effects to the $B \to PP$ decays.
This formalism incorporates the basic elements of
naive factorization approach as its leading term and perturbative QCD
corrections as subleading contributions allowing one to calculate radiative
corrections to the naive factorization for the hadronic two body decays of
$B$ meson.

In our case the total effective Hamiltonian for $B\to PP$ decays
have two sources, the usual SM part $H^{SM}_{eff}$
and the SUSY part $H_{eff}^{susy}$. We use the SM results
obtained in Ref.\cite{BBNS}. The results for the two types of
SUSY contribution from $LL$ and $RR$ mixing can be easily obtained once one
understands the SM contributions since the operators from SUSY interactions
discussed here are all strong penguin type.  In the $LL$ mixing case, the
total WC's are the sum of SM WC ($C^{SM}_{3..6}$) and SUSY contributions.

\begin{eqnarray}
&&C_3^{total} = C_3^{SM} + C^{LL}_3,\;\;\;\;
C_4^{total} = C_4^{SM} + C^{LL}_4,\nonumber\\
&&C_5^{total} = C_5^{SM} + C^{LL}_5,\;\;\;\;
C_6^{total} = C_6^{SM} + C^{LL}_6.
\end{eqnarray}
For $ RR$ mixing, due to the different chirality of the new operators
the total WC's in the final results are replaced by
$C^{total}_{3,4,5,6} = C^{SM}_{3,4,5,6} - C^{RR}_{3,4,5,6}$.
The two type of mixing, $LL$ and $RR$ have similar contributions. Knowing how
one of them affects $B\to K\pi, \pi\pi$ decays, the other can be easily
estimated.

We write down the decay amplitude of $\bar B^0 \to \pi^+\pi^-,  K^- \pi^+ $
for an illustration,

\begin{eqnarray}
% pi+pi- %
A(\bar B^0 \to\pi^{+}\pi^{-})&=&
\frac{G_{F}}{\sqrt{2}}if_{\pi}(m_{B}^{2}-m_{\pi}^{2})F^{B\to\pi}(0)
\Big[ V_{ub}V_{ud}^* (a_{1}+a_{4}^{u}
+a_{10}^{u}  \nonumber \\
&& +R_{\pi}(a_{6}^{u}+a_{8}^{u}))
 +V_{cb}V_{cd}^*( a_{4}^{c}+a_{10}^{c}+R_{\pi}(a_{6}^{c}+a_{8}^{c}
)) \Big] \nonumber \\
 && +i\frac{G_{F}}{\sqrt{2}}f_{B}f^{2}_{\pi}\bigg[V_{ub}V_{ud}^* b_{1}
  +(V_{ub}V_{ud}^*+ V_{cb}V_{cd}^*)
  \Big(b_{3}+2b_{4}-\frac{1}{2}b_{3}^{ew}+\frac{1}{2}b_{4}^{ew}
  \Big)\bigg] \\
%K^-\pi^+%
A(\bar B^0\to K^- \pi^+)&=&{G_F\over \sqrt{2}} if_K(m^2_B-m^2_K)
F_0^{B\to \pi}(m^2_K)
[V_{ub}V_{us}^*(a_1+a_4^u +a_{10}^u+ R_K(a^u_6+a_8^u))\nonumber\\
&&+V_{cb}V_{cs}^*(a_4^c+a_{10}^c + R_K(a^c_6+a_8^c)]\nonumber\\&&
+{G_F\over \sqrt{2}} if_B f_\pi f_K (V_{ub}V_{us}^*+V_{cb}V_{cs}^*)
(b_3 - {1\over 2}b_3^{ew})
\end{eqnarray}
Similarly for other $B\to K \pi, \pi\pi, KK$ modes.

Here $R_K = 2m^2_K/m_s m_b$. $a_i$ and $b_i$ coefficients are related to the
WC's.
Including the lowest $\alpha_s$ order corrections, $a^q_i$'s are given by
\begin{eqnarray}
&&a_1 = C_1 +{C_2\over N}\left[1 + {C_F \alpha_s \over 4\pi} V_K \right]
+\frac{C_2}{N_c}\frac{C_F \pi \alpha_s}{N_c}H_{K\pi} \nonumber\\
&&a_2 = C_2 +{C_1\over N}\left[1 + {C_F \alpha_s \over 4\pi} V_\pi \right]
+\frac{C_1}{N_c}\frac{C_F \pi \alpha_s}{N_c}H_{\pi K} \nonumber\\
&&a_4^p = C^{total}_4 +{C^{total}_3\over N}\left[1 + {C_F \alpha_s \over 4\pi} V_K \right]
+ {C_F \alpha_s \over 4\pi} {P^p_{K,2}\over N_c}+\frac{C^{total}_3}{N_c}
\frac{C_F \pi \alpha_s}{N_c}H_{K\pi } \nonumber\\
&&a_6^p = C^{total}_6 +{C^{total}_5\over N}\left[1 -6\cdot{C_F \alpha_s \over 4\pi} \right]
+ {C_F \alpha_s \over 4\pi} {P^p_{K,3}\over N_c} \nonumber\\
&&a_7 = C_7 +{C_8\over N}\left[1 + {C_F \alpha_s \over 4\pi}
(-V_\pi^{\prime})\right]+\frac{C_8}{N_c}\frac{C_F \pi \alpha_s}{N_c}(-H'_{\pi K}) \nonumber\\
&&a_8^p = C_8 +{C_7\over N}\left[1 -6\cdot {C_F \alpha_s \over 4\pi} \right]
+ {\alpha_{em} \over 9\pi} {P^{p,EW}_{K,3}\over N_c} \nonumber\\
&&a_9 = C_9 +{C_{10}\over N}\left[1 + {C_F \alpha_s \over 4\pi} V_\pi\right]
+\frac{C_{10}}{N_c}\frac{C_F \pi \alpha_s}{N_c}H_{\pi K} \nonumber\\
&&a_{10}^p = C_{10} +{C_9\over N}\left[1 + {C_F \alpha_s \over 4\pi} V_K \right]
+ {\alpha_{em} \over 9\pi} {P^{p,EW}_{K,2}\over N_c}+\frac{C_9}{N_c}\frac{C_F \pi \alpha_s}{N_c}H_{K\pi }
\end{eqnarray}
where $C_F = (N_c^2-1)/(2N_c)$, and $N_c=3$. The quantities $V_M^{(\prime)}$,
$H^{(\prime)}_{M_2 M_1}$, $P^p_{K,2}$, $P^p_{K,3}$, $P^{p,EW}_{K,2}$ and $P^{p,EW}_{K,3}$
are hadronic parameters that contain all nonperturbative dynamics.
The vertex corrections $V_M$ ($M=\pi,K$) are given by
\begin{eqnarray}
V_M &=& 12 \ln{m_b \over \mu} - 18 + \int^1_0 dx g(x) \phi_M(x) \\
V_M^{\prime} &=& 12 \ln{m_b \over \mu} - 6 + \int^1_0 dx g(1-x)\phi_M(x) \\
g(x)&=&  3({1-2x\over 1-x} \ln(x) -i\pi)+\Big[ 2 Li_2(x)-\ln^2(x)+\frac{2\ln(x) }{1-x}
-(3+2i\pi)-(x\leftrightarrow 1-x)\Big].
\end{eqnarray}
where $\phi_M(x)$ is the leading-twist
light cone distribution amplitude of light pseudoscalar meson~\cite{BALL}. This
distribution amplitude can be expanded in Gegenbauer polynomials. We
truncate this expansion at $n=2$.
\begin{eqnarray}
\phi_M(x,\mu) = 6x(1-x)\left[ 1+\alpha_1^M(\mu)C_1^{(3/2)}(2x-1)
+\alpha_2^M(\mu) C_2^{(3/2)(2x-1)}\right]
\end{eqnarray}
where $C_1^{(3/2)}(u)=3u$ and $C_2^{(3/2)}(u)= \frac{3}{2}(5u^2-1)$.
The distribution amplitude parameters $\alpha_{1,2}^M$ for $M=K,\pi$ are:
$\alpha_1^K=0.3$, $\alpha_2^K=0.1$, $\alpha_1^\pi=0$ and $\alpha_2^\pi=0.1$.
The detailed expressions can be found in \cite{BBNS}.

For the penguin contributions,
$P^p_{K,2},\;P^{p}_{K,3},\;P^{p,EW}_{K,2}$ and $P^{p,EW}_{K,3}$ are given by
\begin{eqnarray}
P^p_{K,2}&=&C_1 \left[ \frac{4}{3}\ln \frac{m_b}{\mu}+\frac{2}{3}-G_K(s_p)\right]+
C^{total}_3 \left[ \frac{8}{3}\ln \frac{m_b}{\mu}+\frac{4}{3}-G_K(0)-G_K(1)\right] \nonumber \\
&&+(C^{total}_4+C^{total}_6) \left[ \frac{4 n_f}{3}\ln \frac{m_b}{\mu}-(n_f-2)G_K(0)-G_K(s_c)-G_K(1)\right]\nonumber \\
&&-2C_{11} \int^1_0 \frac{dx}{1-x}\phi_K(x) \nonumber \\
P^{p,EW}_{K,2}&=&(C_1+N_c C_2) \left[ \frac{4}{3}\ln \frac{m_b}{\mu}+\frac{2}{3}
-G_K(s_p)\right]-3C_{12} \int^1_0 \frac{dx}{1-x}\phi_K(x) \nonumber \\
P^{p}_{K,3}&=&C_1 \left[ \frac{4}{3}\ln \frac{m_b}{\mu}+\frac{2}{3}-\hat G_K(s_p)\right]+
C^{total}_3 \left[ \frac{8}{3}\ln \frac{m_b}{\mu}+\frac{4}{3}-\hat G_K(0)-\hat G_K(1)\right] \nonumber \\
&&+(C^{total}_4+C^{total}_6) \left[ \frac{4 n_f}{3}\ln \frac{m_b}{\mu}-(n_f-2)\hat G_K(0)
-\hat G_K(s_c)-\hat G_K(1)\right]-2 C_{11} \nonumber \\
P^{p,EW}_{K,3}&=&(C_1+N_c C_2) \left[ \frac{4}{3}\ln \frac{m_b}{\mu}+\frac{2}{3}
-\hat G_K(s_p)\right]-3C_{12}
\end{eqnarray}
with
\begin{eqnarray}
G_K(s)&=&\int_0^1 dx G_K(s-i\epsilon,1-x)\phi_K(x)  \nonumber\\
\hat G_K(s)&=&\int_0^1 dx G_K(s-i\epsilon,1-x)\phi^K_p(x)  \nonumber\\
G_K(s,x)&=&-4\int_0^1 du u(1-u)\ln [s-u(1-u)x]
\end{eqnarray}
where $\phi_p(x)=1$.

The hard spectator contributions to the coefficients $a_i$'s are parametrized
in terms of a single (complex) quantity $H^{\prime}_{M_1 M_2} $ which suffers
from large theoretical uncertainties related to the regularization of the
divergent endpoint integral. Following \cite{BBNS} we use
$H_{\pi K} = H_{KK}=0.99$ at the scale $\mu = m_b$ and
\begin{eqnarray}
H_{\pi K}^{\prime}=H_{\pi K},\;\;\;\;\;\;\;H_{K\pi}=R_{\pi K}H_{\pi K}.
\end{eqnarray}

All scale-dependent quantities for hard spectator contributions are evaluated
at $\mu_h=\sqrt{\Lambda_h m_b}$ with $\Lambda_h=0.5$ GeV.

Terms containing the coefficients $b_i$ and $b_i^{ew}$ are from annihilation
contributions.
The annihilation coefficients are
\begin{eqnarray}
&&b_1=\frac{C_F}{N_c^2}C_2 A_1^i,\;\;\;\;b_3=\frac{C_F}{N_c^2}
[ C^{total}_3 A^i_1+C^{total}_5 (A^i_3+A^f_3)+N_c C^{total}_6A_3^f] \nonumber \\
&&b_2=\frac{C_F}{N_c^2}C_2 A_1^i,\;\;\;\;b_4=\frac{C_F}{N_c^2}
[ C^{total}_4 A^i_1 +C^{total}_6 A_2^i] \nonumber \\
&&b_3^{ew}=\frac{C_F}{N_c^2} [C_9 A^i_1+C_7(A_3^i+A_3^f)+N_c C_8 A_3^f] \nonumber \\
&&b_4^{ew}=\frac{C_F}{N_c^2}[C_{10} A_1^i+C_8 A^i_2].
\end{eqnarray}
with
\begin{eqnarray}
&&A_1^i=A_2^i=\pi \alpha_s
\left[ 18 \left(X_A-4+\frac{\pi^2}{3} \right)+2r_\chi^2 X_A^2 \right] \nonumber\\
&&A_3^f=12\pi \alpha_s r_\chi^2 (2X_A^2-X_A)
\end{eqnarray}
where $r_\chi=2 \mu_h /m_b$, $X_A=\int^1_0 dy/y$ parameterizes the divergent
endpoint integrals. We use $X_A=\ln(m_B/ \Lambda_h)$.
All scale dependent quantities for annihilation contributions are evaluated at $\mu_h$.

In the QCD improved factorization method the weak
annihilation contributions are power suppressed as $\Lambda_{QCD}/m_b$ and
hence do not appear in the factorization formula. Apart from this, they
also show end-point singularities even at twist-two order in the light-cone
expansion of the final state light mesons and therefore can not be calculated
systematically in the context of hard scattering approach. One can bypass this
problem, by treating these different end-point singularities
as a phenomenological parameters. But this induces model dependence and
numerical uncertainties in the calculation.

Using the above technique we compute the different two-body
branching ratios of $B$ mesons within the framework of SUSY and
compare the calculated results with the experimental data to
constrain the different flavor changing parameters of SUSY. The
experimental branching ratios used in this analysis are listed in
Table~\ref{exp}. We average the data from CLEO, BaBar, Belle
$B$-factories as their results are uncorrelated. To impose the
limit on these flavor changing parameters we require that the
theoretically computed branching ratios should not deviate from
the experimental measurement by $2\sigma$. If the experimental
branching ratio has only upper limit, we select the most stringent
one from the experimental data. It is clear from Table~\ref{exp}
that for $B\to \pi^0\pi^0$ and $B\to KK $ modes have only the
$90\%$ C.L. upper limit.

\begin{table}[htb]
\scriptsize
%\footnotesize
\begin{center}
\caption{The branching ratios for $B\to PP$ in units of $10^{-6}$.}\label{exp}
\begin{tabular}{|l|l|l|l|c|}
 Branching ratio and  & Cleo&Belle& Babar    &Averaged \\
 CP asymmetries       &\cite{4-4}  &\cite{4-5} &\cite{4-6} & Value \\ \hline
$Br(B_u\to \pi^-\bar K^0 )$     &    $18.8^{+3.7+2.1}_{-3.3-1.8}$       &   $22.0\pm 1.9 \pm 1.1$             &   $17.5^{+1.8}_{-1.7}\pm 1.3$                     &         $19.7\pm 1.5$\\   \hline
$Br(B_u\to \pi^0K^-)$           &     $12.9^{+2.4+1.2}_{-2.2-1.1}$      &    $12.8  \pm 1.4 ^{+1.4}_{-1.0}$          &    $12.8^{+1.2}_{-1.1}\pm 1.0$                    &       $12.8\pm 1.1$\\ \hline
$Br(B_d\to \pi^+K^-)$           &    $18.0^{+2.3+1.2}_{-2.1-0.9}$       &    $18.5\pm 1.0\pm 0.7$                   &    $17.9\pm 0.9 \pm0.7$                           &        $18.2 \pm 0.8$\\   \hline
$Br(B_d\to \pi^0\bar K^0)$      &    $12.8^{+4.0+1.7}_{-3.3-1.4}$       &    $12.6 \pm 2.4\pm 1.4$            &     $10.4\pm 1.5\pm 0.8$                          &        $11.2\pm1.4$\\\hline
$Br(B_u\to \pi^-\pi^0)$         &    $4.6^{+1.8+0.6}_{-1.6-0.7}$        &    $5.3\pm 1.3 \pm 0.5$            &    $5.5^{+1.0}_{-0.9}\pm 0.6$                    &         $5.3 \pm 0.8$\\           \hline
$Br(B_d\to \pi^+\pi^-)$         &        $4.5^{+1.4+0.5}_{-1.2-0.4}$    &    $4.4 \pm 0.6 \pm 0.3$                   &    $4.7 \pm 0.6 \pm 0.2$                           &            $4.6\pm 0.4$\\ \hline
$Br(B_d\to\pi^0\pi^0)$          &    $<4.4(90\% \mbox{C.L.})$           &   $<4.4(90\% \mbox{C.L.})$                   &    $<3.6(90\% \mbox{C.L.})$                    &         $<3.6(90\% \mbox{C.L.})$ \\           \hline
$Br(B_u\to K^-K^0)$             &       $<3.3(90\% \mbox{C.L.})$        &    $<3.4(90\% \mbox{C.L.})$             &      $<1.3(90\% \mbox{C.L.})$                 &         $<1.3(90\% \mbox{C.L.})$ \\   \hline
$Br(B_d\to K^-K^+)$             &       $<0.8(90\% \mbox{C.L.})$        &       $<0.7(90\% \mbox{C.L.})$            &      $<0.6(90\% \mbox{C.L.})$                      &  $<0.6(90\% \mbox{C.L.})$ \\  \hline
$Br(B_d\to\bar K^0 K^0)$        &       $<3.3(90\% \mbox{C.L.})$        &      $<3.2(90\% \mbox{C.L.})$            &       $<7.3(90\% \mbox{C.L.})$                    &  $<3.2(90\% \mbox{C.L.})$  \\
\end{tabular}
\end{center}
\end{table}
\normalsize

For our numerical calculations,
we use $m_b(m_b)=4.2~{\rm~GeV}$ for the b quark mass,
$m_c(m_b) = 1.3\pm 0.2$~GeV,
$m_s(2~\rm GeV) = 110.0\pm 0.25$~MeV, $ (m_u+m_d)(2\rm GeV)=
9.1\pm 2.1 $ MeV. For the decay constants and
form factors, we use $f_\pi = 0.131~{\rm~GeV}$,
$f_{K} = 0.160~{\rm~GeV}$, $f_B = 0.180~{\rm~GeV}$, $F^{B\to \pi} = 0.28$,
$r_{\pi K}\simeq \frac{F^{B\to K} f_\pi}{F^{B\to \pi} f_K}=0.9$\cite{BBNS,pdg}.
For the CKM matrix elements, we fix $\lambda = 0.2196$,
and take the central values of $\mid V_{cb}\mid = 0.0412\pm
0.002$, $\mid\frac{V_{ub}}{V_{cb}}\mid = 0.087\pm 0.018$~\cite{pdg}.
We fix the CP violating phase $\gamma$ to $53.6^\circ $ from global fitting
of CKM parameters. With the above input parameters, the SM contributions are
fixed. However, in SUSY the branching ratios depend on several parameters,
such as $(\delta_{j3})_{LL}$, the squark mass $m_{\tilde q}$, and the
ratio $x = m^2_{\tilde g}/m^2_{\tilde q}$. We treat  $m_{\tilde q}$ and
$x$ as free parameters and constrain $(\delta_{j3})_{LL}$
from the available experimental data.
Our main aim will be to obtain limits on the complex plane of
the flavor changing parameters $(\delta_{j3})_{LL}$ for a given $x$ and
the squark mass. Otherwise mentioned, throughout our analysis we will
use 100 GeV for squark mass. Constraints on $(\delta_{j3})_{LL}$ for other
values of $m_{\tilde q}$ can be obtained by rescaling the constraint by
$(m_{\tilde q}/100\mbox{ GeV})^2$.

In the theoretical computations of the branching ratios,
there are many sources of uncertainties, mainly arising from different
form factors, CKM matrix elements, the annihilation contribution
and other hadronic parameters of the QCD improved factorization methods.
We have taken the default values of the above quantities following \cite{BBNS}
to get the central values of the theoretically calculated
branching ratios. However, the results may change if one varies any of the
above parameters within their allowed range.
To take into account this fact we conservatively allow
an extra $50\%$ variation in the theoretically calculated branching ratios
from their default values. In the later section we will discuss more on these
uncertainties. The input values chosen here are for an
illustration. The main issue of our analysis is to show that
$B\to PP$ can provide comparable or even better constraints in
certain area of SUSY parameter space compared with other
well known rare processes for B mesons, like $b\to s \gamma$, $B \to \rho
\gamma$, and the mixing parameters $\Delta m_{B_s}$ and $\Delta m_{B_d}$.

\section{$B\to PP (\Delta S = -1)$, $b\to s\gamma$ and $B^0_s-\bar B^0_s$ mixing}

In this section we study constraints on the parameter $(\delta_{23})_{LL}$
from $B\to K \pi$ and compare with the constraint from $b\to s \gamma$.
We also comment on how $B^0_s -\bar B^0_s$ mixing can provide information
of FCNC parameters in SUSY.
Since we are primarily interested in constraining the flavor changing
parameter $(\delta_{23})_{LL}$, we will present our results on
the complex plane of the $(\delta_{23})_{LL}$ for a given value of
$m_{\tilde q}$ and $x$.

\begin{figure}
\begin{tabular}{cc}
\small{(a) $\bar B^0 \to K^{+} \pi^{-}$} & \small{(b) $B^- \to
K^{-} \pi^{0}$ }     \\ \DESepsf(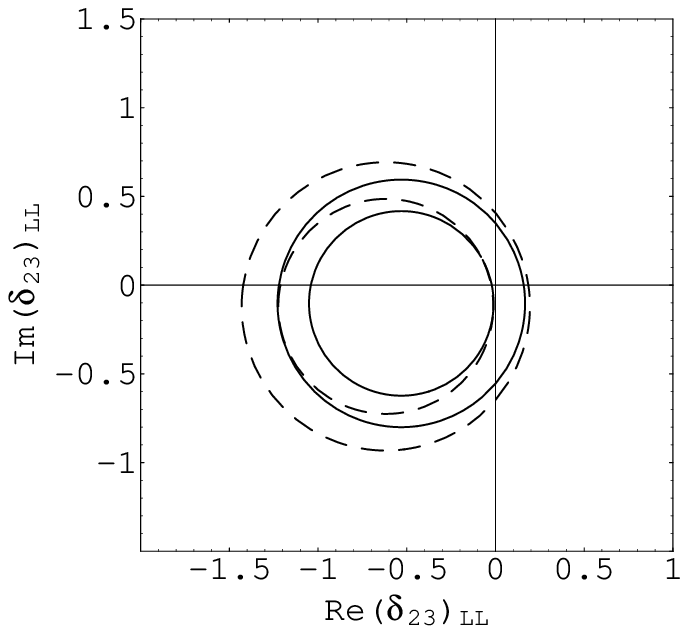 width 7cm)&
\DESepsf(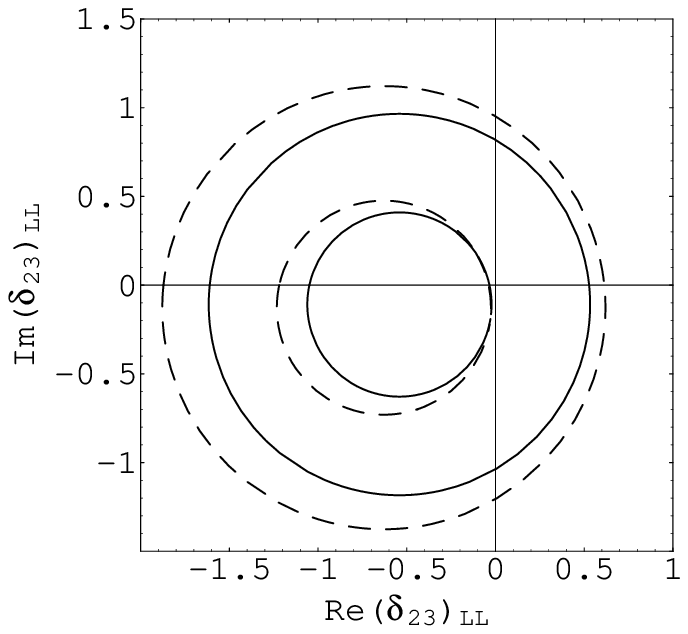 width 7cm)  \\ \small{(c) $B^- \to \bar
K^{0} \pi^{-}$} & \small{(d) $\bar B^0 \to \bar K^{0} \pi^{0}$} \\
\DESepsf(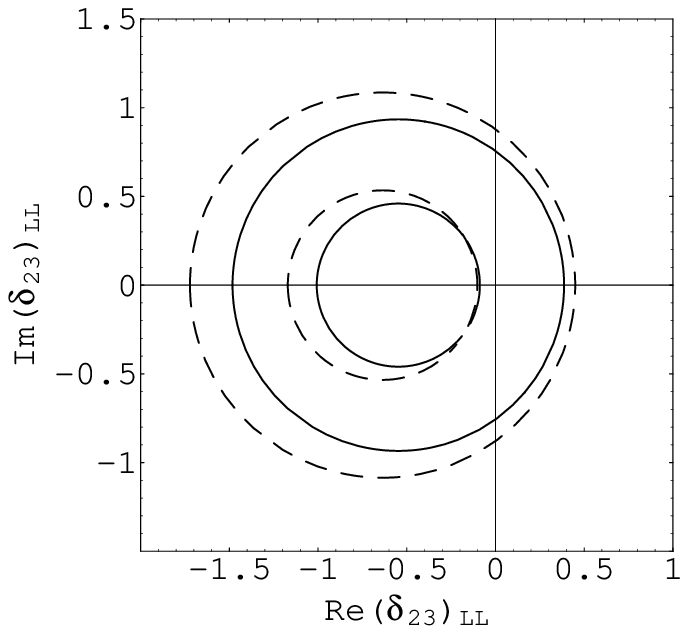 width 7cm) & \DESepsf(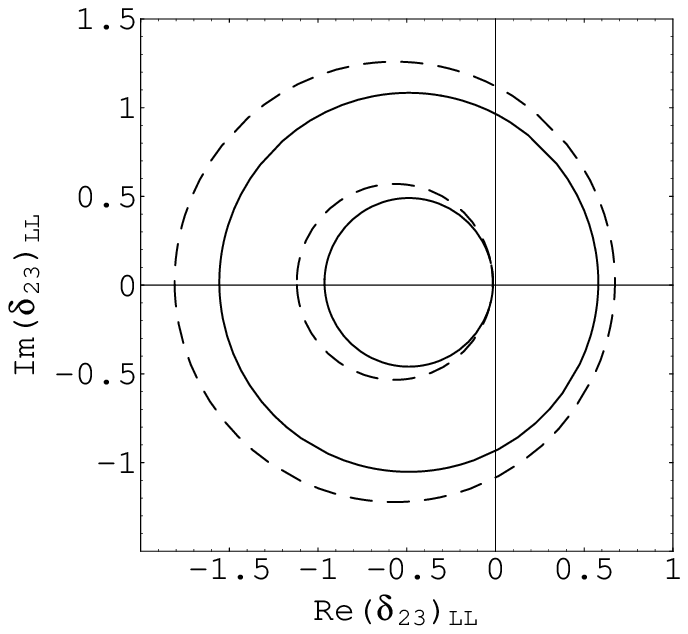 width
7cm)
\\
\end{tabular}
\smallskip
\caption {The $2\sigma$ allowed region on $(\delta_{23})_{LL}$ complex plane
 from $B\to K \pi$ with $m_{\tilde{q}}=100$ GeV. The solid and dashed bands are
the allowed regions
 corresponding to $x=$ 1 and 4 respectively. } \label{dllkpi}
 \end{figure}

In Figure~\ref{dllkpi}  we show $2\sigma $ allowed bands on the
$(\delta_{23})_{LL}$
complex plane for two representative values of $x=1$(solid line) and
4(dashed line) for the squark mass set at $100$ GeV.
From the above Figure with $m_{\tilde{q}}$=100 GeV, it is clear that the four different
decay channels of $B$ into $K^+\pi^-, K^-\pi^0, \bar K^0\pi^-,
\bar K^0 \pi^0 $ give almost similar limits and the allowed ranges are
 almost smaller than 1 on the  $(\delta_{23})_{LL}$ complex plane.
There is no significant change in the obtained bound even when $x$ varies
in the range $0.5\sim 10$.
We finally combine the results obtained from these four branching ratios
and display it in Figure 2. We see that the flavor changing parameter
$(\delta_{23})_{LL}$ is constrained.

If the squark mass $m_{\tilde q}$ is different than the value 100 GeV
used here, one can
obtain the corresponding constraint on $(\delta_{23})_{LL}$ by
multiplying a scaling factor $m^2_{\tilde q}/(100 ~GeV)^2$ on
$(\delta_{23})_{LL}$.
We would like to point out, however, that when making such a rescaling one should
be cautious about the applicability of the mass insertion
method used to obtain the effective Hamiltomian. One should only
apply the results to regions where $(\delta_{23})_{LL}$ is not larger
than of order one. Since the constraint on $(\delta_{23})_{LL}$ is already
of order one for $m_{\tilde q}$
about 100 GeV, for a larger $m_{\tilde{q}}$ $B\to PP$ give weak constraints on
the flavor changing parameters.

\begin{figure}
\center{\DESepsf(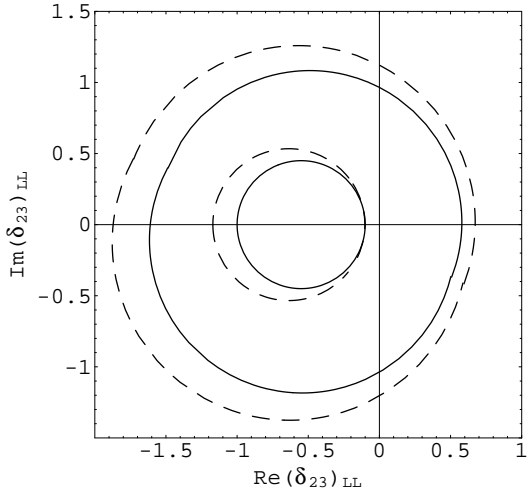 width 7cm)}
\smallskip
\caption {Combined $2 \sigma $ allowed region on
$(\delta_{23})_{LL}$ complex plane obtained from four $ B \to K\pi$ modes as
shown in Figure~\ref{dllkpi}. Other notations are same as Figure~\ref{dllkpi}.}\label{kptotal}
\end{figure}

We now compare the above constraints with
that from $b\to s\gamma$.
We take the current experimental world
average result $Br(B\to X_s \gamma )=(3.41 \pm 0.36)\times 10^{-4}$~\cite{btosgExp}
from CLEO, ALEPH, Belle and Babar. This process is another rare process which
is sensitive to new physics with $\Delta S = -1$ and has been
used to constrain new physics beyond the SM.

The branching ratio of $b\to s \gamma$ is given by~\cite{btosgaLO,btosgaz}
\begin{eqnarray}
Br(B\to X_s \gamma) &=&Br(B \to X e \bar \nu _e) \frac{|V_{ts}^*
V_{tb}|^2}{|V_{cb}|^2} \frac{6 \alpha_{em}}{ \pi g(m_c/m_b) \eta}
|c_{12}(m_b)|^2 \;, \label{btosga}
\end{eqnarray}
where $c_{12}(m_b)=c_{12}^{SM}(m_b)+c_{12}^{susy}(m_b)$,
$g(z)=1-8z^2+8z^6-z^8-24z^4 ln(z)$, and $\eta=1-2f(r,0,0)
\alpha_s(m_b)/3\pi$ with $f(r,0,0)=2.41$\cite{btosgaz}.
The running of the supersymmetric WC's $c^{susy}_{11,12}$ are given
by~\cite{Buras1}
\begin{eqnarray}
c^{susy}_{11}(\mu) = \eta c_{11}^{susy}(M_{susy}),\;\;
c^{susy}_{12}(\mu) = \eta^2 c^{susy}_{12}(M_{susy})+ {8\over
3}(\eta -\eta^2)c^{susy}_{11}(M_{susy}), \label{csusymb}
\end{eqnarray}
where $\eta = (\alpha_s(M_{susy})/\alpha_s(m_t))^{2/21}
(\alpha_s(m_t)/\alpha_s(m_b))^{2/23}$, with

\begin{eqnarray}
&&c_{11}^{susy}(M_{susy}) =\frac{-\alpha_s \pi
(\delta_{j3})_{LL}}{ m_{\tilde q}^2 \sqrt{2}G_F V_{tb}
V_{tq}^*}\Big( -\frac{1}{3} M_3(x)-3 M_4(x)\Big)\;\nonumber\\
&&c_{12}^{susy}(M_{susy}) =\frac{-\alpha_s \pi
(\delta_{j3})_{LL}}{ m_{\tilde q}^2 \sqrt{2}G_F V_{tb}
V_{tq}^*}\Big( -\frac{1}{3}\frac{8}{3} M_3(x)\Big)\;
\end{eqnarray}
The functions $M_{3,4}(x)$ are defined in Ref.\cite{FCNC}.
Here we use the leading order result for $b\to s\gamma $.
The use of the NLO result\cite{btosgaNLO} will change
the LO results by about $10\%$, which will not affect our main conclusion.

\begin{figure}[htb]
\centerline{ \DESepsf(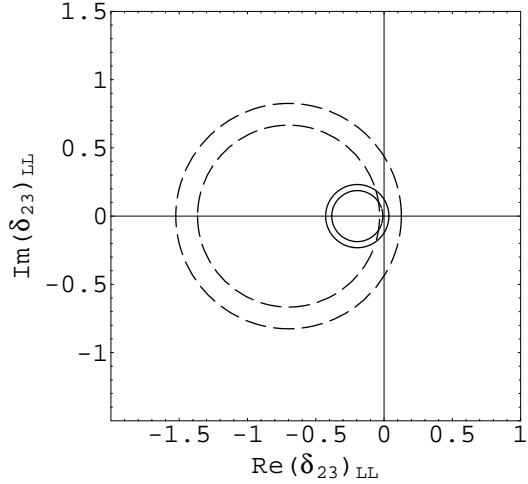 width 7cm)}
\smallskip
\caption {The $2\sigma $ allowed bands obtained from
$b \to s\gamma $ on the complex $(\delta_{23})_{LL}$ plane with $m_{\tilde{q}}=100$ GeV.
Other notations are same as Figure~\ref{dllkpi}.}
\label{plotbtosga}
\end{figure}

The $2\sigma $ constraints on $(\delta_{23})_{LL}$ complex plane
from $b\to s\gamma$ for two different values of $x=1$ and 4 are shown
in Figure \ref{plotbtosga}. From this figure we see that as $x$ increases
the limit on $(\delta_{23})_{LL}$ from $b\to s\gamma$ become weaker.

\begin{figure}
\centerline{\DESepsf(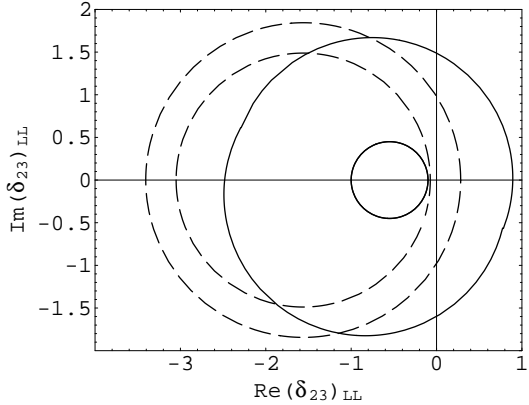 width 7cm)}
    \smallskip
    \caption {Combined $2\sigma$ allowed region on $(\delta_{23})_{LL}$ complex
    plane obtained from  $B\to K\pi$ (solid band) and $b\to s \gamma$ (dashed band)
    for $x=8$ with $m_{\tilde{q}}=100$ GeV.}
    \label{Kpi-btophot}
    \end{figure}

It was thought that $b\to s\gamma$
gives the strongest limit on the parameter $(\delta_{23})_{LL}$.
However, in our analysis we find that in certain region of the parameter
space the bounds obtained from $B \to K \pi$ are stronger than the limit
obtained from $ b\to s\gamma$. From Figure \ref {Kpi-btophot} we see
that for $x = 8 $ constraints from $B \to K \pi$ provide slightly
better constraints than that from $ b\to s\gamma$.
We find that for $x$ less than 8, the constraints obtained
from $B\to PP$ are not as strong as that from $b \to s\gamma$. But for
$x$ larger than 8, the constraints from $B\to PP$ is more stringent.

One can easily understand the above results from the $x$ dependence of
SUSY contributions to $B\to K \pi$ and $b\to s\gamma$.
From  Eq. 7, we find that the SUSY contributions
mainly appear in parameters $a_4$ and $a_6$ through $C^{total}_{3,4,5,6}$.

\begin{eqnarray}
a^{susy} = a^{susy}_4 + a^{susy}_6 =
C^{LL}_4 + {C_3^{LL}\over N} + R_{\pi, K} ( C_6^{LL} + {C_5^{LL}\over N}).
\end{eqnarray}
We have checked numerically that the
other contributions in Eq. 7 from SUSY interactions are small.

\begin{figure}[htb]
\centerline{ \DESepsf(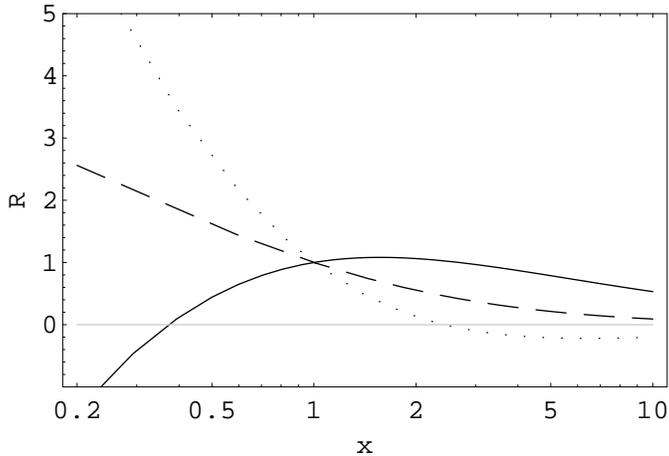 width 10cm)}
\smallskip
\caption {$R(PP)$, $R(q\gamma)$ and $R(B^0_{d,s}-\bar B^0_{d,s})$ as
functions of $x$. The solid, dashed and dotted lines correspond to
$R(PP)$, $R(q\gamma)$, and $R(B^0_{d,s}-\bar B^0_{d,s})$, respectively.}
\label{constkpi}
\end{figure}

The parameter $a^{susy}$
linearly depends on $(\delta_{j3})_{LL}/m^2_{\tilde q}$ and
depends on $x$ in a complicated but known form which we indicate as
$a^{susy} = a^{susy}(x)$.
One can easily study how experimental data may constrain the parameter
$(\delta_{j3})_{LL}/m^2_{\tilde q}$ as a function of $x$ by studying the
behavior of

\begin{eqnarray}
R(PP) = {a^{susy}(x)\over a^{susy}(1)}.
\end{eqnarray}
The above function shows the behavior of the SUSY contribution
with $x$ normalized with the value at $x=1$.
In Figure \ref{constkpi}, we show
the behavior of $R(PP)$ as a function of $x$ by the dark solid line. From the
figure we see that $R(PP)$ has a zero value at $x\simeq0.37$.
This implies that for $x\simeq 0.37$, no constraint on $(\delta_{23})_{LL}$
from $B\to PP$.
In the neighborhood of this region $b\to s\gamma$ certainly give a better
bound. For $x$ in the range $1 \sim 10 $, $R(PP)$ varies very slowly with $x$.
This almost flat behavior of $R(PP)$ makes the bound on $(\delta_{23})_{LL}$
to remain nearly unchanged within this range of $x$.

While for $b\to s \gamma$, the SUSY contribution has a different
$x$ dependence. In Figure \ref{constkpi} we show the variation of
\begin{eqnarray}
R(q\gamma) = {c^{susy}_{12}(x)\over c^{susy}_{12}(1)},
\end{eqnarray}
as a function of $x$ by
the dashed line. We see that $R(q\gamma)$ decreases when $x$ increases.
This means, for a given $m_{\tilde q}$, the constraint on
$(\delta_{23})_{LL}$ becomes weaker as $x$ increases. At certain value of
$x$, $B\to K \pi$ give better bounds than $b\to s\gamma$.
Numerically we find that when $x$ is larger than 8,
$B\to K \pi$ give better constraints on $(\delta_{23})_{LL}$
than $b\to s\gamma$.

We now comment on the implications of $q-\tilde q-\tilde g$ interactions
on $B^0_s -\bar B^0_s$ mixing. The effective Hamiltonian
for this process is given by~\cite{FCNC}
\begin{eqnarray}
&&H_{eff}=-C_m O_m\;,\\ && C_m=\frac{\alpha_s^2}{216 m_{\tilde
q}^2}[ 24x f_6(x)+66\tilde{f}_6 (x)] (\delta_{j3})^2_{LL}\;, \\ &&
O_m=\bar s^\alpha_L \gamma_\mu b^\alpha_L \bar s^\beta_L
\gamma_\mu b^\beta_L
\end{eqnarray}
The functions $f_6(x)$ and $\tilde f_6(x)$ are given by:
\begin{eqnarray}
&&f_6(x)=\frac{6(1+3x)\ln(x) +x^3-9x^2-9x+17}{6(1-x)^5}\;,\\ &&
\tilde f_6(x)=\frac{6x(1+x)\ln(x) -x^3-9x^2+9x+1}{3(1-x)^3}\;.
\end{eqnarray}

After RG evolution we get the WC at $m_b$ scale \cite{runWCBs}:
\begin{eqnarray}
&&C_m(m_b)=\eta_1 C_m(M_{susy})\;,\\ &&\eta_1=\Big(
\frac{\alpha_s(m_t)}{\alpha_s(m_b)}\Big)^{6/23}\Big(
\frac{\alpha_s(M_{susy})}{\alpha_s(m_t)}\Big)^{6/21}\;,
\end{eqnarray}
the total $\Delta M_{B_s}$ is given by $\Delta
M_{B_s}$=$2|M^{SM}_{12}+M^{susy}_{12}|$ with
$M^{susy}_{12}=\eta_1(C_m/3)M_{B_s}B_{B_s} f_{B_s}^2$. The SM
contribution $M^{SM}_{12}$ is given by \cite{Buras2}, and
$M^{susy}_{12}$ contains phase from parameter
$(\delta_{j3})^2_{LL}$. At present we have only stringent lower
bound on $\Delta M_{B_s}$ which is $ 14.4\;ps^{-1}$ at the 95\%
C.L.\cite{bsmix}. Using previously obtained constraints one can
obtain the allowed range for $\Delta M_{B_s}$ with SUSY
contributions. In our numerical calculation we use\cite{pdg}
$\xi\equiv \sqrt{B_{B_s}}/f_{B_s}\sqrt{B_{B_d}} f_{B_d}=1.16\pm
0.05$ and $\sqrt{B_{B_d}} f_{B_d}=230\pm 40MeV$.

\begin{figure}[htb]
\centerline{ \DESepsf(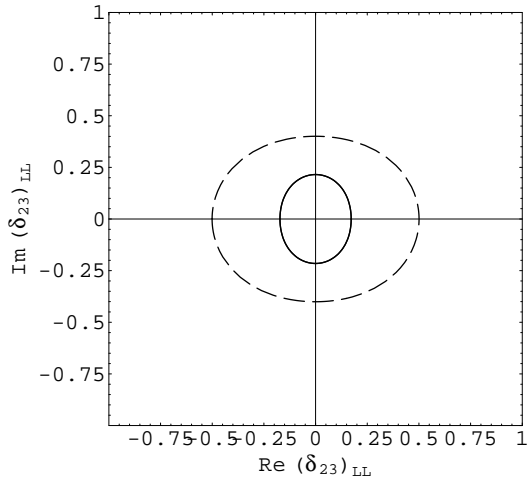 width 7cm)}
\smallskip
\caption {The allowed areas (areas inside each curve are excluded if
$\Delta M_{B_s}$ is larger than $48 \;ps^{-1}$) with
solid and dashed boundaries corresponding to $x = 1$ and 4 respectively
on the complex plane of $(\delta_{23})_{LL}$.}
\label{plotdmbs}
\end{figure}

We find that $B\to K \pi$ and $b\to s \gamma$
data do not constrain $\Delta M_{B_s}$
significantly in a large part of the parameter space allowing
$\Delta M_{B_s}$ to be much larger than the present lower bound. If
future experiments can measure $\Delta M_{B_s}$ of the order or less than
$\sim 48\;ps^{-1}$, which is the limit LHCb can reach, then, one would obtain
a much stronger limit on $(\delta_{23})_{LL}$ for $x = 1$ and 4
as shown in Figure~\ref{plotdmbs}.
However we still find that there are regions where $B\to K \pi$ and
$b\to s \gamma$ give better constraints. To see in which
area of parameter space, the SUSY contributions to $\Delta M_{B_s}$
become negligible , we study the behavior of relative SUSY contribution,
$R(B^0_s-\bar B^0_s)= C_m(x)/C_m(1)$ as a function of $x$. In
Figure~\ref{constkpi} we show the variation of $R(B^0_s-\bar B^0_s)$ as
a function of $x$ by the dotted lines.
we see that at $x = 2.43$ SUSY contribution to $\Delta M_{B_s}$
becomes zero and no constraint on $(\delta_{23})_{LL}$ can be obtained.
In regions around $x= 2.43$, $B\to K\pi$ and $b\to s \gamma$ can give
better constraints even if we assume that the future experiment will
measure a $\Delta M_{B_s}$ less than  $ 48\;ps^{-1}$.

\section{$B\to PP (\Delta S = 0)$, $B\to \rho \gamma$ and $B_d -\bar B_d$
mixing}

\begin{figure}
\begin{center}
\begin{tabular}{cc}
\small{(a) $\bar B^0 \to \pi^{+} \pi^{-}$} & \small{(b) $\bar B^0 \to \pi^0 \pi^{0}$ }     \\
\DESepsf(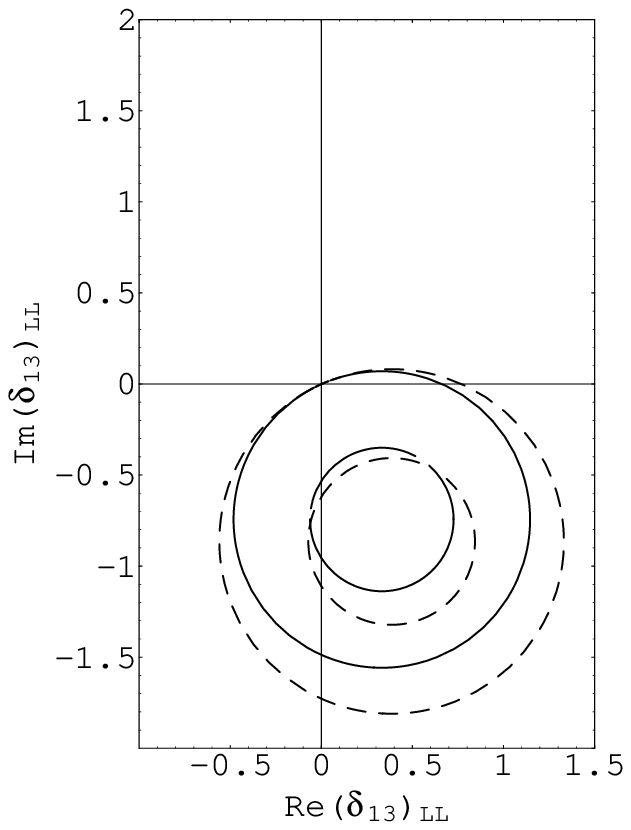 width 5cm)&\DESepsf(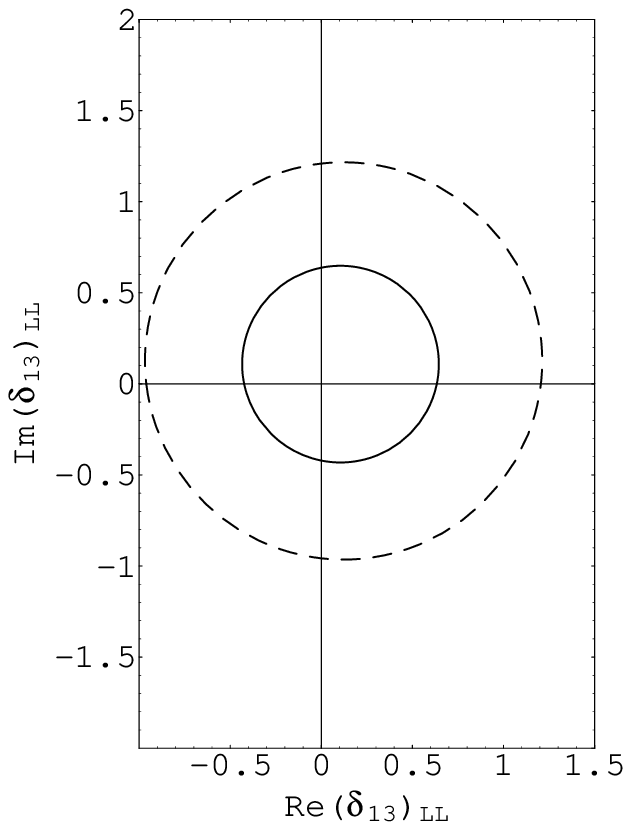 width 5cm)  \\
\small{(c) $B^- \to K^-K^0$} &\small{(d) $\bar B^0 \to K^0 \bar K^{0} $} \\
\DESepsf(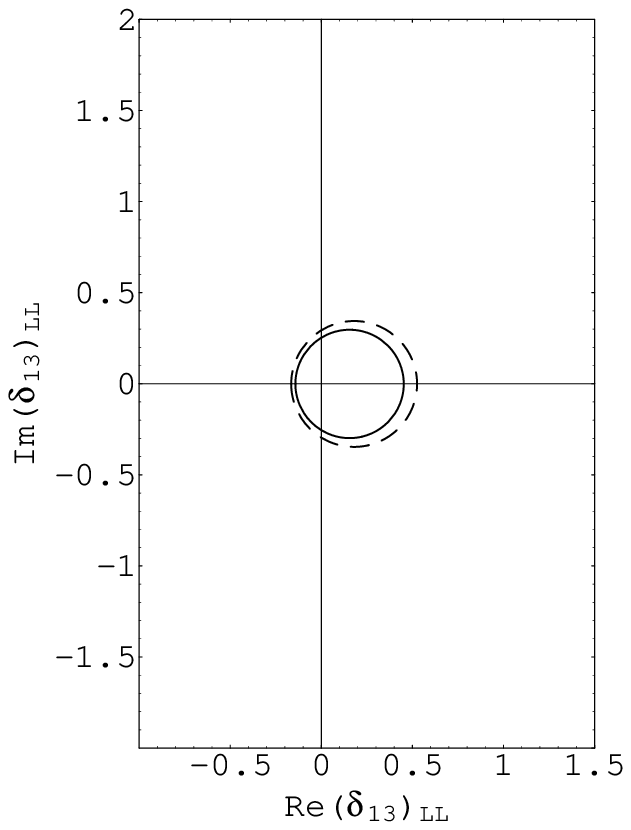 width 5cm) & \DESepsf(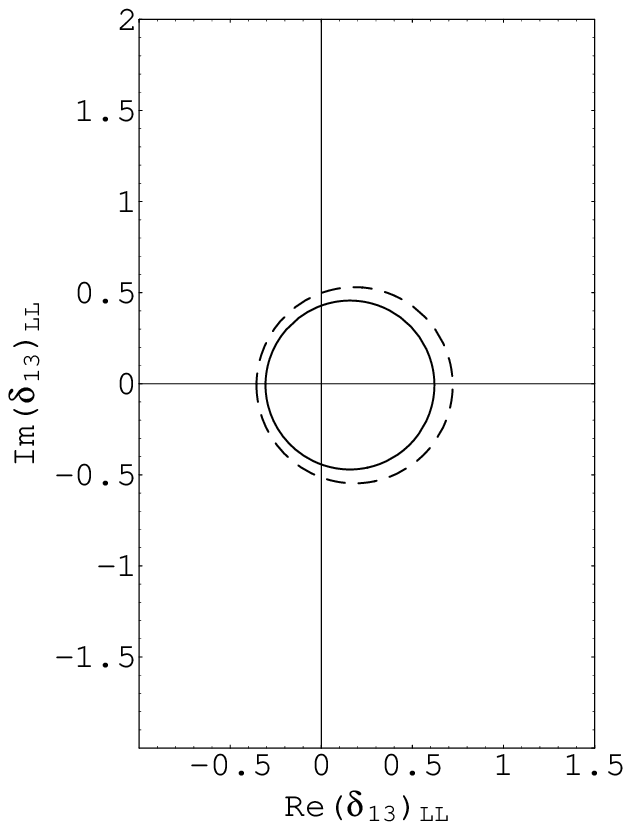 width 5cm)\\
\end{tabular}
\end{center}
\smallskip
\caption {The constraint on $(\delta_{13})_{LL}$ from $B\to \pi
\pi$, $B\to K \bar K$.
For $B\to \pi^+ \pi^-$ are $2\sigma$ allowed region
while for $B\to \pi^0\pi^0$, $B\to K^- K^0$ and $B\to K^0 \bar K^0$ is
the limits with $90\%$C.L. The solid and dashed bands (a) and
circles ((b)-(d)) are the allowed regions corresponding to $x=$ 1 and 4
respectively.}
\label{dllpipikk}
\end{figure}

In this section we discuss constraints on SUSY interactions from following
processes: $B \to \pi \pi, K K$, $B \to \rho \gamma$
and $\Delta M_{B_d}$.
Constraints obtained from $B\to \pi\pi$ and $B\to K K$
are shown in Figure~\ref{dllpipikk}. We find that the constraints on
$(\delta_{13})_{LL}$ are better than $(\delta_{23})_{LL}$ obtained
from $B\to K\pi$ decays. The strongest one is obtained from $B^-\to
K^- K^0$.

\begin{figure}[htb]
\centerline{ \DESepsf(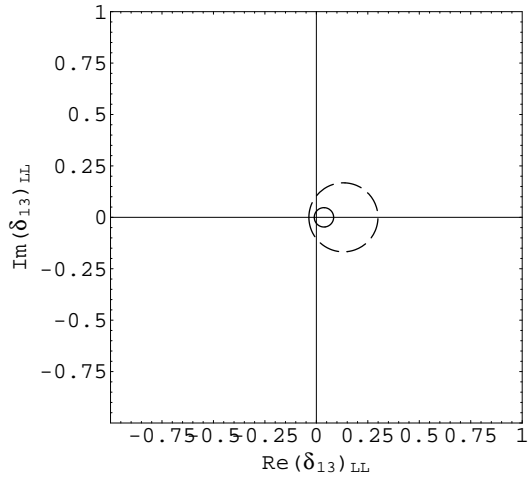 width 7cm)}
\smallskip
\caption {The constraint of $(\delta_{13})_{LL}$ from $B\to \rho \gamma$.
The allowed area marked by solid and dashed line correspond to
$x=1~{\rm and }~4$ respectively.}
\label{rhogamma}
\end{figure}

Although $B\to \rho \gamma$ has not been measured, the experimental
upper bound from BaBar $Br(B^0 \to \rho^0 \gamma)\leq 1.4 \times 10^{-6}$ at
the $90\%$C.L. \cite{btorhogamma} has been
used to obtain a constraint on $(\delta_{13})_{LL}$.
The decay width $\Gamma (B \to \rho \gamma)$ is given by:
\begin{eqnarray}
\Gamma(B \to \rho \gamma)=\frac{\alpha_{em}}{32\pi^4}G_F^2
|V_{tb}|^2 |V_{td}^*| |F^{B\to \rho}(0)|^2 |c_{12}(m_b)|^2
(m_b^2+m_d^2)\frac{(m_B^2-m_\rho^2)^3}{m_B^3}
\end{eqnarray}
We use the form factor calculated by T.~Huang {\it et al.} in
Ref.~\cite{formrho},
$F^{B\to \rho ^0}=0.237\pm 0.035$.
The constraints on $(\delta_{13})_{LL}$ for
$x=1$ and $4$ are shown in Figure~\ref{rhogamma}.
Comparing the constraints obtained from $B \to \pi \pi, K K$
with $B \to \rho \gamma$ we find that similar to the discussions in
the previous section, when $x$ becomes larger the constraint from
$B^-\to K^- K^0$ become relatively better than $B \to \rho \gamma$,
numerically for the strongest case $B^- \to K^- K^0$,
this occurs at $x$ about 6.

%\newpage
%\newpage
\begin{figure}[htb]
\centerline{ \DESepsf(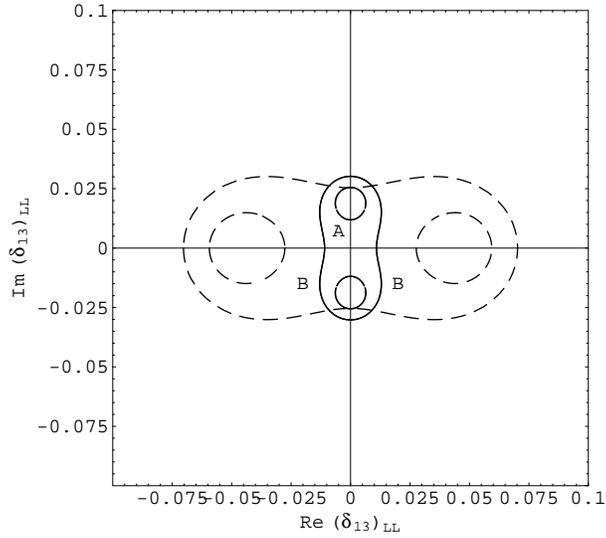 width 8cm)}
\smallskip
\caption {
The areas marked by ``A'' and ``B'' on the complex $(\delta_{23})_{LL}$
plane represent $2\sigma$ allowed regions for $x=1$ and 4 respectively
obtained from $B_d - \bar B_d$ mixing.}
\label{plotdmbd}
\end{figure}
%\newpage

We now turn to constraint from $\Delta M_{B_d}$. This quantity has been
experimentally well measured with
the world average value\cite{pdg} $0.489\pm 0.008\;ps^{-1}$.
However, the theoretical calculation has
uncertainty from $\sqrt{B_d} f_{B_d}$, which is $230\pm 40$ MeV.
The results are shown in
Figure \ref{plotdmbd}. It can be seen that
$\Delta M_{B_d}$ provides a much stronger constraint on $(\delta_{13})_{LL}$
compared with the constraint on $(\delta_{23})_{LL}$ from
$\Delta M_{B_s}$. In a large part of the parameter space
the branching ratios $B\to \pi\pi$, $B\to K \bar K$ provide
weaker constraints on $(\delta_{13})_{LL}$ unless $x$ is
close to 2.43 where SUSY contribute to $\Delta M_{B_d}$ is zero as can be seen
from Figure \ref{constkpi}.
In the region around $x=2.43$
$B\to K^- K^0$ and $B\to \rho \gamma$ give better bounds.

\section{Constraints on $(\delta_{ij})_{RR}$}

In the previous discussions, we have concentrated on constraints
on $(\delta_{ij})_{LL}$. We now discuss constraints on
$(\delta_{ij})_{RR}$. Due to the different chirality of the new
operators the total WC's, in the theoretical expressions for $B\to PP$,
are replaced by
$C^{total}_{3,4,5,6} = C^{SM}_{3,4,5,6} - C^{RR}_{3,4,5,6}$, where
$C^{RR}_{3,4,5,6}$ have the same form as for $C^{LL}_{3,4,5,6}$ in
Eq. 3 with $(\delta_{ij})_{LL}$ replaced by
$(\delta_{ij})_{RR}$.

\begin{figure}[htb]
\centerline{ \DESepsf(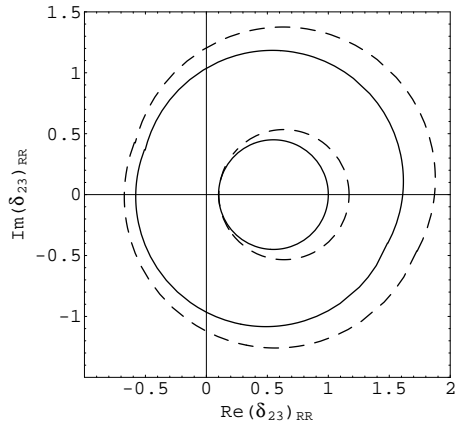 width 6cm)}
\smallskip
\caption {Combined ($\bar B^0 \to K^+\pi^-,\;\bar K^0\pi^0$) and ($B^- \to
K^-\pi^0,\;\bar K^0\pi^-$) constraints on
$(\delta_{23})_{RR}$ with $m_{\tilde{q}}=100$ GeV.}\label{RR-1}
\end{figure}

In Figure \ref{RR-1}, we present the
combined ($\bar B^0 \to K^+\pi^-,\;\bar K^0\pi^0$) and ($B^- \to
K^-\pi^0,\;\bar K^0\pi^-$) constraints on
$(\delta_{23})_{RR}$. The allowed ranges are the same but as a
mirror reflection in complex plane of $(\delta_{ij})_{RR}$
compared with that of $(\delta_{ij})_{LL}$.

\begin{figure}[htb]
\centerline{ \DESepsf(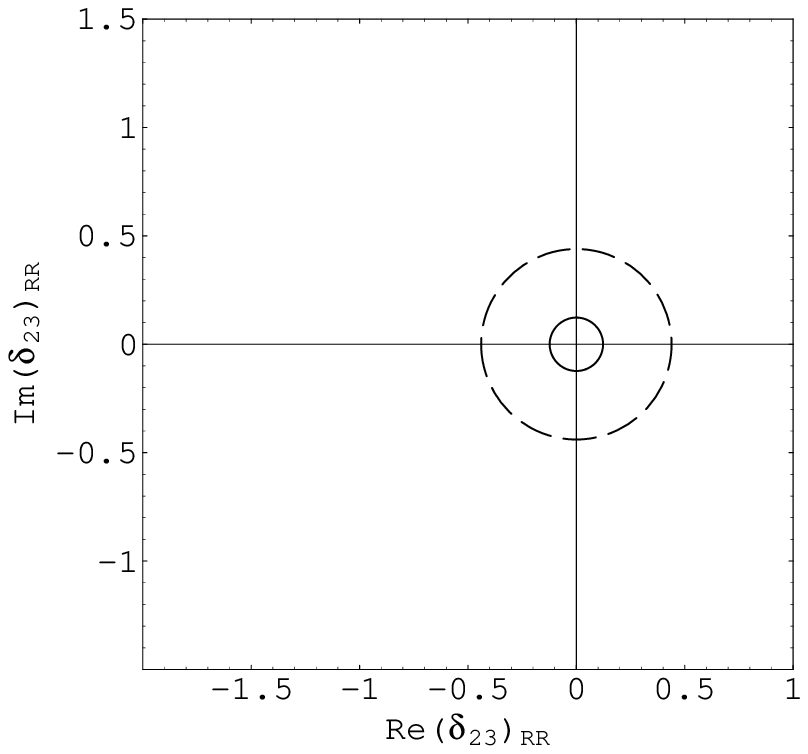 width 6cm) \DESepsf(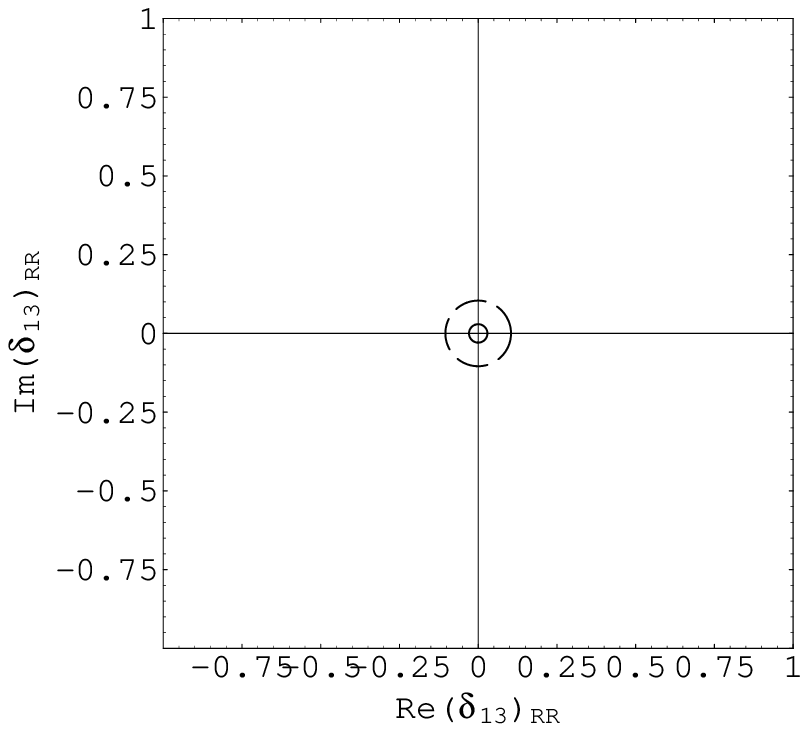 width 6cm)}
\smallskip
\caption {$b \to s\gamma$ and $B \to \rho\gamma$ on
$(\delta_{23})_{RR}$ with $m_{\tilde{q}}=100$ GeV.}\label{RR-2}
\end{figure}

In $b \to s\gamma$ and $B \to \rho\gamma$ cases, because of
different chirality in operators, the Wilson coefficient
$c^{susy}_{12}$ is separated from the SM contribution
$c^{SM}_{12}$. One needs to replace $|c_{12}(m_b)|^2$ in Eq. 18
and Eq. 31 by $|c_{12}(m_b)|^2=|c_{12}^{SM}|^2+|c_{12}^{susy}|^2$.
The allowed ranges are shown in Figure \ref{RR-2}. We see that
they are different than those shown in Figures 3 and 8. It is
clear that the allowed ranges of $(\delta_{ij})_{RR}$ are
constrained to be smaller than $(\delta_{ij})_{LL}$.

$B_{d,s}-\bar B_{d,s}$ mixing also constrain $(\delta_{ij})_{RR}$.
In this case one just needs to replace $(\delta_{ij})_{LL}$ in Eq. 25
by $(\delta_{ij})_{RR}$ and $Q_m=\bar s^\alpha_L \gamma_\mu b^\alpha_L \bar s^\beta_L
\gamma_\mu b^\beta_L$ by $\tilde{Q}_m=\bar s^\alpha_R
\gamma_\mu b^\alpha_R \bar s^\beta_R \gamma_\mu
b^\beta_R$. Since $\langle B^0|Q_m|\bar B^0\rangle=\langle B^0|{\tilde{Q}}_m|\bar
B^0\rangle$, the constraints on $(\delta_{ij})_{RR}$ from $\Delta m_{B_{d,s}}$
are the same as in Figures 6 and 9.

\section{Conclusions}

In this paper we studied the constraints on the SUSY flavor
changing parameters $(\delta_{13})_{LL, RR}$ and $(\delta_{23})_{LL,RR}$
from rare hadronic $B\to PP$ decays.
We improved the calculations by two folds.
Firstly, all the relevant Wilson coefficients
are computed at $M_{susy}$ and then evolved down to lower scale $\mu = m_b$
using Renormalization Group Equation.
Secondly, we calculated the
two body hadronic $B$ decays using QCD improved factorization method.
We found that the values of the
Wilson coefficients do change significantly under this evolution which in
turn affect the theoretical computation of $B$ decays.
The constraints obtained are compared with radiative
$b\to s \gamma$ and $B\to \rho \gamma$ decays, and $\Delta M_{B_{d,s}}$.

We found that each of the different processes discussed can provide
interesting constraints on the SUSY flavor changing parameters within
certain parameter space. For $(\delta_{23})_{LL,RR}$ we found that as long as
$x \leq 7$, $b \to s\gamma $ provides the most stringent limit, however,
for $x \geq 8$, $B \to K\pi$ give better constraints.
For $(\delta_{13})_{LL,RR}$ we observed $\Delta M_{B_d}$ provides a better
bound, but for $x$ around 2.43, the bound becomes weaker. In this region,
$B\to K^- K^0$ and $B\to \rho \gamma$ can give stronger constraint.

Before we conclude, we would like to comment on the possible theoretical
uncertainties arising from different form factors, CKM elements, the hard
spectator interactions and also from annihilation contributions in QCD improved
factorization scheme. As we have already mentioned earlier that these
hard spectator interactions and annihilation contributions in QCD improved
factorization method suffer from endpoint singularities and these singularities
have been treated as a phenomenological parameters inducing model dependence
and additional source of numerical uncertainties. However, the exact
estimations of these uncertainties are not yet known. The uncertainties
in the different form factors
can change the theoretical results by about $40\%$. We think that the overall
uncertainty could be around $50\%$. So, in our theoretical calculations we have
taken into account this $50\%$ uncertainty to obtain the limits.
In our calculations we also fixed the SM phase $\gamma$ to be
$53.6^\circ$. When SUSY contributions are included, the $\gamma$ phase will
differ from its SM value. Allowing $\gamma$ to vary will also
change the details of the numerical results.

However our main objective here was to explore all possible $B\to PP$ decays
within the context of SUSY and show how in different regions of parameter space
the different decay modes can constrain the FCNC parameters.  The generic
feature of our analysis will not change significantly by the above mentioned
different uncertainties.
A coherent study of different processes involving $B$-mesons
carried out here can serve to provide detailed information about the
SUSY FCNC parameters. We hope in future with the improved theoretical
calculations and data will provide a better understanding of
SUSY flavor changing interactions.

\acknowledgements
%\section{Acknowledgements}
This work is partially supported by the ROC National Science Council
under the grant NSC 90-2811-M-002-054, NSC 90-2112-002-058
and by the ROC Ministry of Education
Academic Excellence Project 89-N-FA01-1-4-3.

%....................................................................%
\def\pr#1, #2 #3 { {\em Phys. Rev.}         {\bf #1},  #2 (19#3)}
\def\prd#1, #2 #3{ {\em Phys. Rev.}        {D \bf #1}, #2 (19#3)}
\def\pprd#1, #2 #3{ {\em Phys. Rev.}       {D \bf #1}, #2 (20#3)}
\def\prl#1, #2 #3{ {\em Phys. Rev. Lett.}   {\bf #1},  #2 (19#3)}
\def\pprl#1, #2 #3{ {\em Phys. Rev. Lett.}   {\bf #1},  #2 (20#3)}
\def\plb#1, #2 #3{ {\em Phys. Lett.}        {\bf B#1}, #2 (19#3)}
\def\pplb#1, #2 #3{ {\em Phys. Lett.}        {\bf B#1}, #2 (20#3)}
\def\npb#1, #2 #3{ {\em Nucl. Phys.}        {\bf B#1}, #2 (19#3)}
\def\pnpb#1, #2 #3{ {\em Nucl. Phys.}        {\bf B#1}, #2 (20#3)}
\def\prp#1, #2 #3{ {\em Phys. Rep.}        {\bf #1},  #2 (19#3)}
\def\zpc#1, #2 #3{ {\em Z. Phys.}           {\bf C#1}, #2 (19#3)}
\def\epj#1, #2 #3{ {\em Eur. Phys. J.}      {\bf C#1}, #2 (19#3)}
\def\mpl#1, #2 #3{ {\em Mod. Phys. Lett.}   {\bf A#1}, #2 (19#3)}
\def\ijmp#1, #2 #3{{\em Int. J. Mod. Phys.} {\bf A#1}, #2 (19#3)}
\def\ptp#1, #2 #3{ {\em Prog. Theor. Phys.} {\bf #1},  #2 (19#3)}
\def\jhep#1, #2 #3{ {\em J. High Energy Phys.} {\bf #1}, #2 (19#3)}
\def\pjhep#1, #2 #3{ {\em J. High Energy Phys.} {\bf #1}, #2 (20#3)}
\def\epj#1, #2 #3{ {\em Eur. Phys. J.}        {\bf C#1}, #2 (19#3)}
\def\eepj#1, #2 #3{ {\em Eur. Phys. J.}        {\bf C#1}, #2 (20#3)}
%....................................................................%

\end{document}